\documentstyle[aps,prb,epsfig,multicol]{revtex}
\begin{document}
\title{Kinetic Monte Carlo simulations of the growth of polymer crystals}
\author{Jonathan P.~K.~Doye and Daan Frenkel}
\address{FOM Institute for Atomic and Molecular Physics, Kruislaan 407,\\ 
1098 SJ Amsterdam, The Netherlands}
\date{\today}
\maketitle
\begin{abstract}
Based upon kinetic Monte Carlo simulations of crystallization in
a simple polymer model we present a new picture of the mechanism by which 
the thickness of lamellar polymer crystals is constrained to a value close 
to the minimum thermodynamically stable thickness, $l_{\rm min}$.
The free energetic costs of the polymer extending beyond the edges of
the previous crystalline layer and of a stem being shorter than $l_{\rm min}$
provide upper and lower constraints on the length of stems in a new layer.
Their combined effect is to cause
the crystal thickness to converge dynamically to a value close to $l_{\rm min}$ where
growth with constant thickness then occurs.
This description contrasts with those given by the two dominant theoretical approaches.
However, at small supercoolings the rounding of the crystal profile does inhibit 
growth as suggested in Sadler and Gilmer's {\em entropic barrier} model.
\end{abstract}

\begin{multicols}{2}
\section{Introduction}

On crystallization from solution and the melt many polymers form lamellae
where the polymer chain traverses the thin dimension of the crystal many times folding back
on itself at each surface.\cite{Keller57a}
Although lamellar crystals were first observed over forty years ago
their physical origin is still controversial.
It is agreed that the kinetics of crystallization are crucial since extended-chain crystals are
thermodynamically more stable than lamellae.
However, the explanations for the dependence of the lamellar thickness on temperature
offered by the two dominant theoretical approaches appear irreconcilable. \cite{theorynoteb,alttheory}
The lamellar thickness is always slightly greater than $l_{\rm min}$, the minimum thickness for which
the crystal is thermodynamically more stable than the melt;
$l_{\rm min}$ is approximately inversely proportional to the degree of supercooling.

The first theory, which was formulated by Lauritzen and Hoffman (LH) soon after the initial
discovery of the chain-folded crystals,\cite{Lauritzen60,Hoffman76a,Hoffman97}
invokes {\em surface nucleation} of a new layer on the thin side faces of the lamellae as the key process.
It assumes that there is an ensemble of crystals of different thickness,
each of which grows with constant thickness.
The crystals which grow most rapidly dominate this ensemble,
and so the average value of the thickness in the ensemble, which is equated with the observed thickness,
is slightly larger than the thickness for which the crystals have the maximum growth rate.
The growth rates are derived by assuming that a new crystalline layer grows by the deposition
of a succession of stems (straight portions of the polymer that traverse the crystal once)
along the growth face.
The two main factors that determine the growth rate are the thermodynamic driving force
and the free energy barrier to deposition of the first stem in a layer.
The former only favours crystallization when the thickness is greater than $l_{\rm min}$;
the latter increases with the thickness of the crystal because of the free energetic cost
of creating the two new lateral surfaces on either side of the stem and makes
crystallization of thick crystals prohibitively slow.
Therefore, the growth rate passes through a maximum at an intermediate value of the thickness
which is slightly greater than $l_{\rm min}$.

The second approach, which was developed by Sadler and Gilmer and has been termed the {\em entropic barrier} model,
is based upon the interpretation of kinetic Monte Carlo simulations \cite{Sadler84a,Spinner95}
and rate-theory calculations\cite{Sadler86a,Sadler87d,Sadler88a}
of a simplified model of polymer crystal growth. As with the surface nucleation approach,
the observed thickness is suggested to result from the competition between a driving force
and a free energy barrier contribution to the growth rate.
However, a different cause for the free energy barrier is postulated.
As the polymer surface in the model can be rough,
it is concluded that the details of surface nucleation of new layers are not important.
Instead, the outer layer of the crystal is found to be thinner than in the bulk;
this rounded crystal profile prevents further crystallization.\cite{Sadler88a}
Therefore, growth of a new layer can only begin once a fluctuation occurs to an entropically
unlikely configuration in which the crystal profile is `squared-off'.
As this fluctuation becomes more unlikely with increasing crystal thickness,
the entropic barrier to crystallization increases with thickness.

Although both approaches are able to describe correctly some of the basic
phenomenology of polymer crystallization, both have questionable aspects.
As Frank and Tosi pointed out, one implication of the LH assumptions is that 
the thickness of an individual crystal should not vary even when the 
temperature is changed.\cite{Frank60}
However, this is clearly contradicted by experiment:
when the temperature is changed the thickness of a growing crystal adjusts 
to the new temperature producing a step on the lamella.\cite{Bassett62,Dosiere86a}
Sadler has also extensively questioned the adequacy of the LH approach, in particular
to explain curved crystal habits and the effect of twins on the growth 
rate.\cite{Sadler83a,Sadler84b,Sadler86b,Sadler87a,Sadler87b}
Furthermore, there is the so-called `$\delta l$-catastrophe':
at too large supercoolings the predicted thickness goes
to infinity unless a parameter in the model, the apportionment factor $\Psi$, 
is rather arbitrarily set to zero.\cite{Hoffman97}

In the entropic barrier approach it is not clear whether approximations such as
the implicit representation of the chain connectivity and chain folds
by a set of simple growth rules and the neglect of the energetic contribution of chain 
folds to the free energy of the fold surface compromise its conclusions.
Furthermore, although the simulations have been interpreted in terms of an entropic barrier,
no direct evidence of this free energy barrier has been provided.\cite{notfree}

There have been a number of works which have taken the basic LH approach
but relaxed some of the constraints. 
For example, both Frank and Tosi\cite{Frank60} and 
Lauritzen and Passaglia\cite{Lauritzen67b} have allowed the stem 
lengths within a layer and between layers to vary.
Later, Point allowed the first stem to be deposited in steps rather than as 
a complete stem; this change prevents the $\delta l$-catastrophe.\cite{Point79a,Point79b}
In the same spirit DiMarzio and Guttman wanted to allow every stem to be deposited in
steps and to allow a fold to be formed at any step; 
however they could only consider very restricted cases.\cite{DiMarzio82a}
The approach used in all these papers was to attempt to 
find analytical solutions to the rate-theory equations, but
as the number of possible pathways considered increased 
the problems became increasingly intractable.
The natural solution to this difficulty
is to use computer simulation techniques such as kinetic Monte Carlo (KMC).
However, all these papers are over fifteen years old, 
and at that time such an approach was not so feasible; 
only Point made an attempt to apply computational methods to the problem.\cite{Point79c}

Therefore, a thorough application of computer simulation to 
an unrestricted version of the LH approach is long overdue.
In this paper we do just this. 
We examine the growth of new crystalline layers from solution on a 
surface that represents the growth face of a polymer crystal when there are no 
constraints on the stem length or on when a fold can form. 
Our model is, in some ways, midway between the original LH model and the model
used by Sadler and Gilmer. In the latter, a realistic representation of the polymer is 
sacrificed in order to be able to consider, for example, rough growth surfaces.
Consequently, our work, as well as providing a test of the two theories, 
could play an important role in understanding the relationship between them.
This work should also be seen in the context of an increasing application of
simulation to study processes relevant to polymer crystallization.\cite{Yamamoto97,Chen98,Toma98}

\section{Methods}
In our simulations the polymer is represented by a self-avoiding walk on a simple cubic lattice. 
There is an attractive energy, $-\epsilon$, between non-bonded polymer units on 
adjacent lattice sites and an energetic penalty, $\epsilon_g$, for each kink (a `gauche bond') in the chain. 
As the surface in our system represents the growth face of a polymer crystal the $-\epsilon$
interaction also applies to polymer units in contact with the surface.
$\epsilon$ can be considered to be an effective interaction representing 
the combined effects of polymer-polymer, polymer-solvent and solvent-solvent interactions,
and so we can use our model in a simplified representation of the crystallization 
of a semi-flexible polymer from solution.
The behaviour of the polymer is controlled by the ratio $kT/\epsilon$; 
large values can be considered as either high temperature or good solvent conditions, 
and low values as low temperature or bad solvent conditions.
The parameter $\epsilon_g$ defines the stiffness of the chain. The polymer
chain is flexible at $\epsilon_g$=0 and becomes stiffer as $\epsilon_g$ increases.
Here, we use $\epsilon_g=8\epsilon$,
however similar results are obtained at any positive $\epsilon_g$.\cite{stiff}

This interaction scheme has been recently used to investigate the phase behaviour of isolated
homopolymers\cite{Doniach96a,Bastolla97a,Doye98a} and a homopolymer in the presence of 
a surface.\cite{Doye98c}
The low temperature behaviour is of particular interest since the polymers were found to adopt 
cuboidal\cite{Doye98a} (rectangular when on a surface\cite{Doye98c})
`crystalline' configurations involving chain folding.
However, these structures were observed for thermodynamic reasons---they 
are lowest in energy---whereas chain-folding occurs in polymer crystallization 
for kinetic reasons.

\begin{figure}
\begin{center}
\epsfig{figure=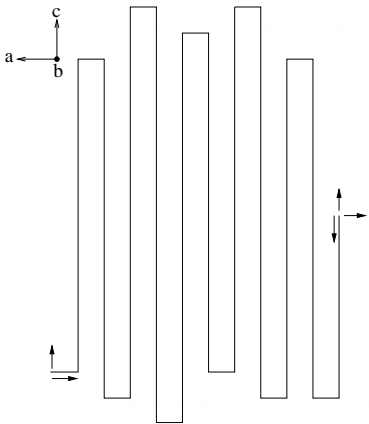,width=5cm}
\end{center}
\vspace{3mm}
\begin{minipage}{8.5cm}
\caption{\label{fig:moves}An example configuration for the crystalline portion
of the polymer which shows the possible moves (represented by arrows) 
in the kinetic Monte Carlo simulation. 
For this configuration, where growth is allowed to occur from either end,
there are five possible moves. The dashed lines mark the boundaries of the 
underlying surface
}
\end{minipage}
\end{figure}

In our simulations we wish to investigate the growth of 
new crystalline layers on the growth face of a crystal.
To achieve this we choose not to use conventional Monte Carlo or molecular dynamics simulations
since it is not feasible with these methods to probe the desired time and length scales.
Instead we use kinetic Monte Carlo.\cite{Voter86}
However, this, of course, means that we have to make certain assumptions about the processes 
contributing to polymer crystallization.
In our model when a polymer crystallizes on the growth face it forms a 
two-dimensional crystalline layer.
The crystals formed only have adjacent re-entry of the stems; i.e.\ only tight folds where the
new stem is adjacent to the previous stem are allowed. 
An example configuration is shown in Figure \ref{fig:moves}.

To make the problem more computationally tractable
we only model the crystalline portion of the polymer explicitly. 
The rest of the chain is assumed to behave like an ideal coil. 
If we consider the coil to be a two-dimensional coil adsorbed onto the surface the free energy is
\begin{equation}
A_{\rm ideal,2D}=-N_{\rm coil} k T\log\left( 1+ 2 \exp(-\beta \epsilon_g)\right) - N_{\rm coil} \epsilon,
\label{aid.2D}
\end{equation}
whereas if the coil adopts a three-dimensional configuration in solution the free energy is
\begin{equation}
A_{\rm ideal,3D}=-N_{\rm coil} k T\log\left( 1+ 4 \exp(-\beta \epsilon_g)\right),
\label{aid.3D}
\end{equation}
where $N_{\rm coil}$ is the number of units in the coil state.
These expressions ignore any energetic contributions from contacts between 
polymer units in the coil. 
This approximation becomes worse as the temperature decreases, 
because the disordered chain would then be expected to have a dense collapsed conformation.\cite{Doye98a}

For a single homopolymer (with the current interactions scheme
and with all units explicitly represented) in the presence of 
an infinite polymer-like surface, the lowest energy configuration is a two-dimensional 
crystalline configuration. 
Simulations have shown that on melting of the crystalline polymer a 
two-dimensional disordered state was formed for all values of $\epsilon_g$,
and only at higher temperature does the polymer become more three-dimensional.\cite{Doye98c}
This suggests that Equation (\ref{aid.2D}) is an appropriate form to describe the coil.
However, when considering growth on a thin lamella, 
the surface area on which the coil can absorb is much reduced. 
To stay on the surface the coil must
adopt an anisotropic configuration in the plane which would reduce its entropy
and thus reduce the free energetic advantage of a 2D coil compared to a 3D coil.
Although we predominantly use Equation (\ref{aid.2D}) to describe the coil, we also
consider the effects of assuming the coil is 3D.

At each step changes of configuration can only occur at the ends of the 
crystalline portion of the polymer. 
The possible processes are for the crystalline part of the polymer to grow by one unit 
by the extension of one of the stems or by the formation of a gauche bond as part of a fold, or for the crystalline
part of the polymer to shrink by one unit. 
To preserve adjacent re-entry after the formation of the first of the two gauche bonds 
of a fold the possible moves are just the completion of the fold and the removal of the first
gauche bond.
These possibilities are illustrated in Figure \ref{fig:moves}.
Generally we allow growth to occur from both ends of the crystalline configuration,
but occasionally, in order to compare with theories of polymer crystallization, 
we only allow growth to occur from one end. 
Physically, these two scenarios correspond to the initial 
crystal nucleus being in the mid-section of the polymer, or at the end of the polymer,
respectively. We see no reason why the latter should always be the case. 
The choice only makes a difference to the growth rate, 
which is of course twice as fast when growth can occur from both ends, 
and the structure of the crystal near to the initial nucleus.

Note that our scheme does not allow for the annealing of the part of the crystallite 
that has already been deposited. This will not affect our conclusions concerning the 
average thickness of the deposited layers but it may lead to an unrealistically rough
appearance of the crystal. We will come back to this issue later in the paper.

The change in the free energy of the polymer on taking a step, $\Delta A$, is given by
$\Delta A=\Delta E_{\rm xtal}+\Delta A_{\rm coil}$, where $\Delta E_{\rm xtal}$ is the change
in energy of the crystalline configuration and $\Delta A_{\rm coil}$ is calculated from 
Equation (\ref{aid.2D}) or (\ref{aid.3D}).
A rate is assigned to each of the possible moves. Each move is assumed to be an activated
process with a barrier that is an energy $\Delta$ above the higher free energy state. 
The prefactor is assumed to be the same for all processes, and we set it to 1 thus
defining a time unit.
Therefore, the rate for a step is given by 
\begin{equation}
k_{ij}=\exp\left( {-\Delta\over kT}\right) \min\left( 1,\exp \left({-\Delta A_{ij}\over kT}\right) \right).
\end{equation}
As the factor $\exp\left( -\Delta/kT\right)$ scales all the rates, simulations do not need to be carried
out at different values of $\Delta$, but the results for one value can be scaled onto any other. 

In the KMC simulation at each step we choose randomly one state, $j$, from those 
connected to the current state, $i$, with a probability given by
\begin{equation}
P_{ij}={k_{ij}\over\sum_{j'}k_{ij'}},
\end{equation}
and update the time by an increment 
\begin{equation}
\Delta t= -{\log(\rho)\over\sum_{j}k_{ij}},
\end{equation}
where $\rho$ is a random number in the range [0,1]. 
In this way the KMC algorithm simulates a stochastic process 
described by a Poisson distribution.

As the expressions for the various rates do not depend on the length of the polymer
we do not 
set a limit to the length of the polymer, but allow
each new crystalline layer to be formed from a single polymer.

Most of the results in this paper are for the growth of a single crystalline layer on 
a growth face which is of uniform thickness. 
However, we have also considered the situation where many layers are grown on the initial crystal.
In these simulations all except the first layer are grown on top of the previously-grown layer 
and we only start to grow a new layer once the previous layer has been completed (when
the crystalline polymer spans the periodic boundary conditions in the $a$-direction).

\end{multicols}
\begin{figure}
\begin{center}
\epsfig{figure=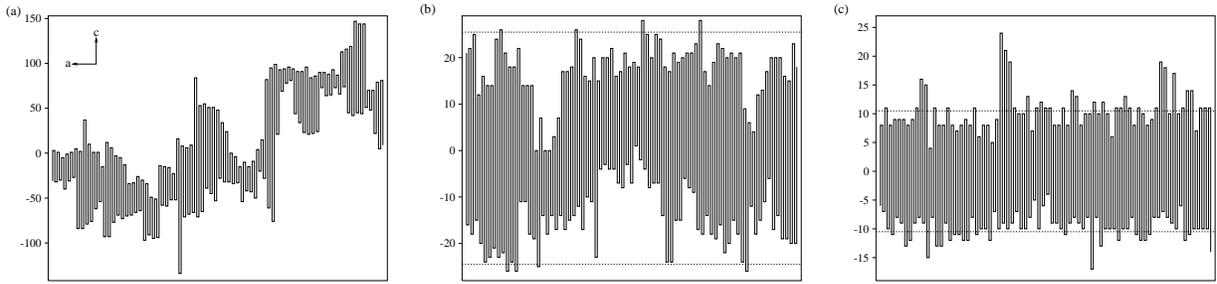,width=16.5cm}
\vglue0.1cm
\begin{minipage}{17.6cm}
\caption{\label{fig:config}
Polymer configurations for a new crystalline layer grown at $T=2.75\,\epsilon k^{-1}$ 
on a surface of thickness: (a) $\infty$, (b) 50 and (c) 21. 
Each configuration contains 150 stems and a dashed line marks the boundaries of 
the underlying surface; the $y$-axis origin is at the centre of the surface.}
\end{minipage}
\end{center}
\end{figure}
\begin{multicols}{2}

\begin{figure}
\begin{center}
\epsfig{figure=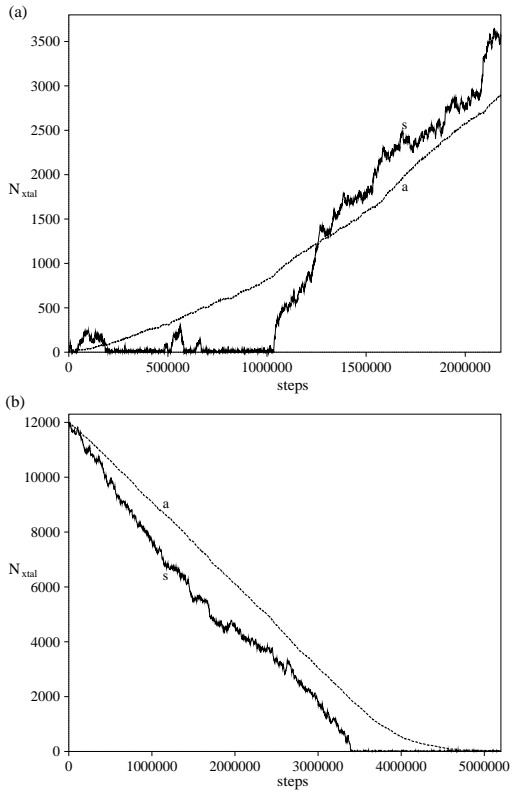,width=7cm}
\end{center}
\vspace{1mm}
\begin{minipage}{8.5cm}
\caption{\label{fig:grow}
The number of units in the crystalline layer, $N_{\rm xtal}$, as a function
of the number of steps for (a) $T=3.7\,\epsilon k^{-1}$,
(b) $T=4.1\,\epsilon k^{-1}$. The solid line (s) shows the behaviour of a single layer, whereas the
dashed line (a) is an average over 50 layers. The underlying surface is infinite. Initially, there is 
in (a) a single unit on the surface and in (b) a crystal 100 stems wide, each 120 units long.
}
\end{minipage}
\end{figure}

The thermodynamics of this model can be easily worked out if we ignore 
any entropic contribution to the crystal from variations in the stem length.
The energy of a crystal with constant thickness, $l$, is 
\begin{equation}
E_{\rm xtal}= N_{\rm xtal} \left({2\epsilon_g+\epsilon\over l} -2\epsilon\right),
\end{equation}
where $N_{\rm xtal}$ is the number of units in the crystalline configuration.
The positive term is due to the surface energy of the fold surface. 
The minimum stable thickness, $l_{\rm min}$, can be 
calculated through the equation $E_{\rm xtal}/N_{\rm xtal}=A_{\rm coil}/N_{\rm coil}$.
This gives 
\begin{equation}
l_{\rm min,2D}={2\epsilon_g+\epsilon\over
                \epsilon-k T\log\left( 1+ 2 \exp(-\beta \epsilon_g)\right)}
\label{eq:lmin.2D}
\end{equation}
and
\begin{equation}
l_{\rm min,3D}={2\epsilon_g+\epsilon\over
                2\epsilon-k T\log\left( 1+ 4 \exp(-\beta \epsilon_g)\right)}.
\label{eq:lmin.3D}
\end{equation}
The neglect of the entropy of the crystal means that these expressions are upper bounds.
The temperature at which the surface of the polymer crystal loses its crystalline order, 
the `melting point' $T_m$, is given by the temperature for which the denominator is zero.

\section{Results}

Some aspects of the typical behaviour of our model for the growth of a single layer on a surface
are illustrates in Figures \ref{fig:config} and \ref{fig:grow}.
Below the melting point ($T_m=4.06\,\epsilon k^{-1}$), for a sufficiently thick surface, 
a new crystalline layer grows. However, at low supercoolings
the initial nucleation of a new layer can be quite slow. 
In the example shown in Figure \ref{fig:grow}a, there is no net growth in the first million 
time steps, and only once a viable nucleus has formed does growth occur with relatively little impediment. 
This example clearly shows the effect on the dynamics of the free energy barrier for the 
initial nucleation of a layer. 
The time scale for the initial nucleation event decreases with increasing supercooling
and at large supercoolings it is not noticeable in the trajectories.
(Very close to $T_m$ it is only feasible to grow new layers when a seed crystal is introduced.)

Above $T_m$, as one would expect, if we start the simulation with an initial 
seed crystal it duly dissolves. In the example shown in Figure \ref{fig:grow}b the temperature
is just above $T_m$, and so the rate of dissolution is slow 
because of the nearly-flat free energy landscape. In every 1000 steps there is only 3 more 
dissolution than growth steps.

In the LH theory the thickness of a layer is determined by the length 
of the first stem that is deposited. 
However, as the free energy barrier associated with 
the first stem is proportional to the length of the stem, one could
imagine that growth would be more rapid for a layer where the stem length gradually increases
to its average value as crystallization progresses than for a layer with constant stem length.
In such a configuration the lateral surface free energy is paid for `in installments' rather than 
all initially and growth is likely to be more rapid because of the lower initial free energy barrier.

\begin{figure}
\begin{center}
\epsfig{figure=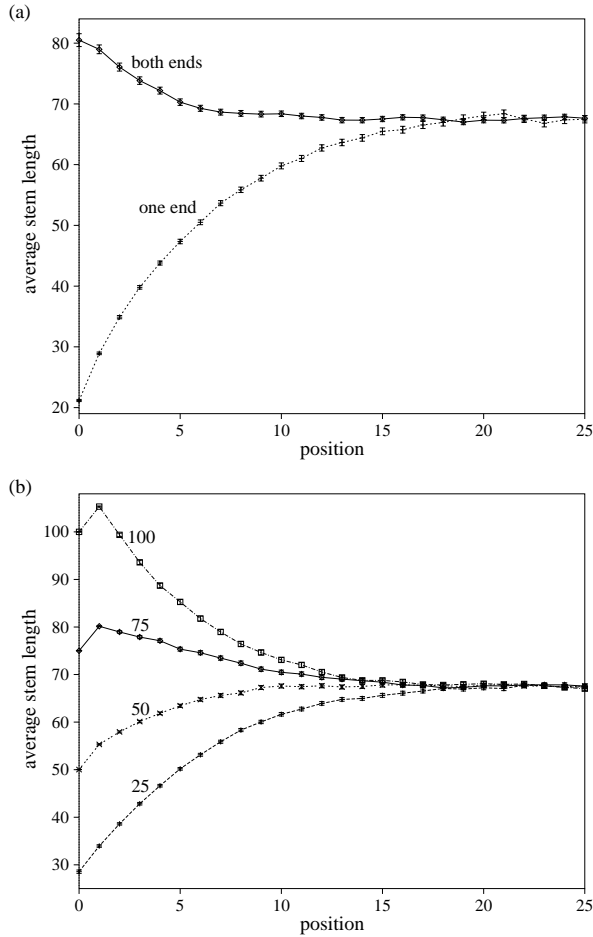,width=8.2cm}
\end{center}
\vspace{3mm}
\begin{minipage}{8.5cm}
\caption{\label{fig:seed}The dependence of the average stem length on the distance of the stem from 
(a) the initial nucleation site, and 
(b) the centre of an initial crystal seed at $T=2.75\,\epsilon k^{-1}$ 
for growth on an infinite surface. 
In (a) growth starts with a single polymer unit on the surface and
we consider the cases where growth is allowed at one end or both ends of the crystalline
portion of the polymer.
In (b) the crystal seeds are 3 stems wide; the lengths of the stems in the seeds are 
as labelled. 
}
\end{minipage}
\end{figure}

Our model allows us to examine these questions easily.
Firstly, inspection of example configurations shows that there 
is considerable variation in the stem lengths within a layer (Figure \ref{fig:config}). 
There is nothing in the kinetics of growth which causes the stem length to be constant. 
(Although subsequent annealing of the new crystalline layer may reduce the roughness.)
Secondly, we have tested whether there is any systematic dependence of the average stem length
on the position of the stem with respect to the initial nucleation site (Figure \ref{fig:seed}).
Allowing the crystal to grow from only one end most closely mimics the conditions of the 
LH theory in which stems are grown one at a time.
In this case we see that the suggestion we made above is confirmed. 
The length of the first stem is significantly shorter than the average stem length in the layer,
and the stem length converges to the average value for the layer as the crystal grows.

Interestingly when we allow growth to occur from both ends 
the initial behaviour is very different (Figure \ref{fig:seed}a). 
This is because it is now possible to form an initial nucleus which consists of 
two stems connected by a fold which grow simultaneously, a scenario not considered in the LH theory. 
Such a nucleus is energetically favoured because of the interactions between the two stems. 
As it avoids the large free energy barrier associated with a single-stem nucleus\cite{Doye98c}
there is now no advantage in the initial stems being shorter than the average in the layer (Figure \ref{fig:seed}a).
The possibility of a two-stem nucleus was first suggested by Point\cite{Point79b}
and recently strong evidence for this nucleus has been obtained in a study of 
the free energy profiles along specific crystallization pathways.\cite{Doye98c} 

\begin{figure}
\begin{center}
\epsfig{figure=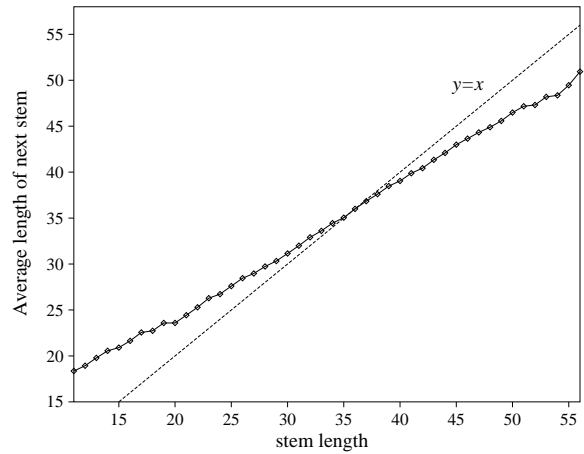,width=8.2cm}
\end{center}
\vspace{3mm}
\begin{minipage}{8.5cm}
\caption{\label{fig:inlayer}
The dependence of the stem length on the length of the previous stem for the 
growth of a single layer on a surface 50 units thick at 
$T=2.75\,\epsilon k^{-1}$. 
}
\end{minipage}
\end{figure}

This finding about the structure of the initial nucleus significantly dents the LH theory because 
its assumption about the nature of the initial nucleus, and 
the consequent free energy barrier, is crucial to the theory.
However, this finding is rendered somewhat irrelevant because 
the above results also undermine 
a more fundamental tenet of the LH theory, 
namely the assumption that the thickness of a layer is determined by the initial nucleus.
Further confirmation that this assumption does not hold comes when we examine the 
growth from initial seed crystals.
Whatever the thickness of the initial seed the thickness of the growing crystal 
converges to the same value (Figure \ref{fig:seed}b).  
These results imply that the thickness of a crystalline layer must be determined
by factors which are operating on the deposition of each stem and 
not those specific to the initial stems. 
We shall examine what these factors are in detail later.

In Figure \ref{fig:inlayer} we show the correlations between successively grown stems.
The figure has the form of a fixed-point attractor.
It shows there is one stem length, $l^{*}$, for which the average length of the next stem is the 
same as the previous (for Figure \ref{fig:inlayer} $l^{*}$=36), 
and this corresponds to the average stem length in the layer. 
For stems longer (shorter) than $l^*$ the length of the next stem is on average shorter (longer)
bringing the stem length nearer to $l^*$.
Frank and Tosi have previously drawn a similar conclusion from their extension of the LH approach
in which a single change in the stem length is allowed during the deposition of a new layer.\cite{Frank60}

\begin{figure}
\begin{center}
\epsfig{figure=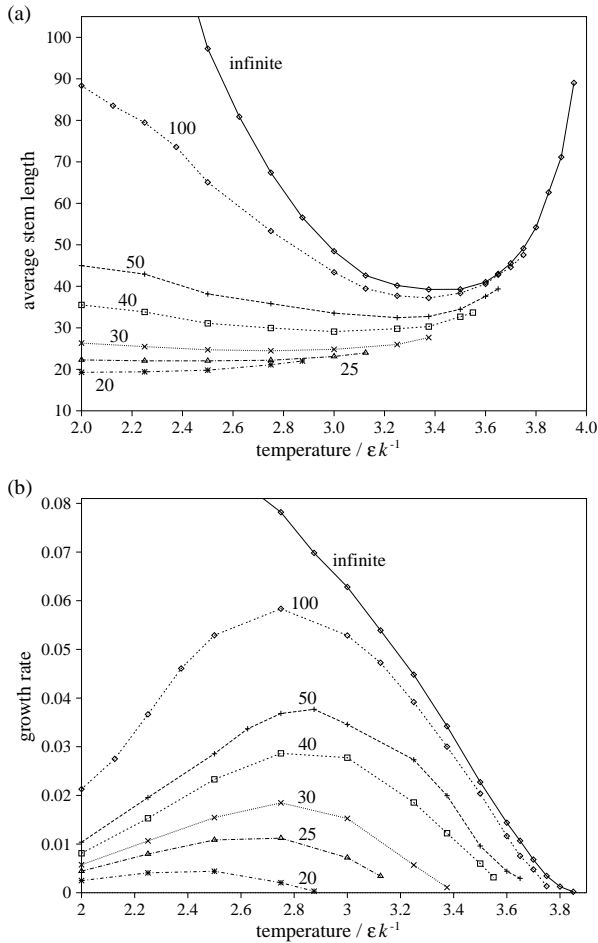,width=8.2cm}
\vglue 0.1cm
\begin{minipage}{8.5cm}
\caption{\label{fig:thick.T} (a) The average stem length and (b) the growth rate 
(polymer units per $\exp(-\Delta/kT$) units of time) in the new crystalline layer as a function of temperature. 
The different curves are for different thicknesses of the underlying surface, as labelled.
}
\end{minipage}
\end{center}
\end{figure}

\begin{figure}
\begin{center}
\epsfig{figure=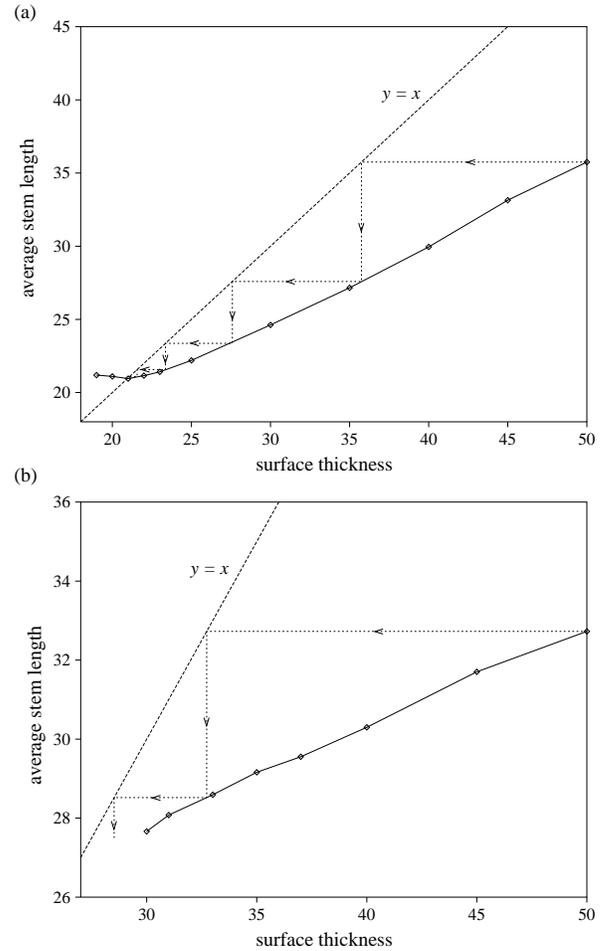,width=8.2cm}
\vglue 0.1cm
\begin{minipage}{8.5cm}
\caption{\label{fig:thick.l} 
The average stem length in the new crystalline layer 
as a function of the thickness of the underlying surface for 
(a) $T=2.75\,\epsilon k^{-1}$, (b) $T=3.375\,\epsilon k^{-1}$.
The dotted lines show how the thickness changes on addition of successive
layers to a 50-unit thick surface.
}
\end{minipage}
\end{center}
\end{figure}

In Figure \ref{fig:thick.T}a we show how the thickness of a new layer depends on the temperature and
the thickness of the growth surface.
It can be immediately seen from the fact that the lines are not horizontal that the thickness 
of a new layer is not necessarily the same as the previous layer, thus contradicting another 
of the assumptions of the LH theory.
Instead the temperature dependence of the curves reflects the factors that determine the thickness of a new layer.
All the thickness curves rise as the temperature approaches $T_m$ because of the rise of $l_{\rm min}$;
they end at the temperature where it is no longer possible to grow a new layer which is thermodynamically stable.
(Figure \ref{fig:thick.T}b shows that the growth rate of the crystal goes to zero at these end points.)
At low temperature the thickness also increases, in this instance, because it becomes increasingly difficult to scale the 
free energy barrier for forming a fold\cite{Doye98c} and so on average the stem continues to grow for longer.
However, except for growth on an infinite surface, this rise is checked by the presence 
of the edge of the underlying surface. 
It is unfavourable for the polymer to overhang the edge because these units do not gain the $-\epsilon$ 
energy of interaction with the surface.

\begin{figure}
\begin{center}
\epsfig{figure=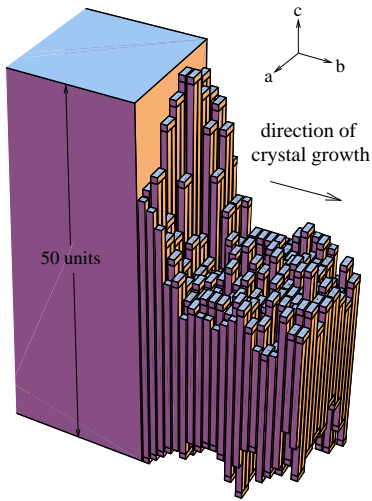,width=7cm}
\vglue -0.2cm
\begin{minipage}{8.5cm}
\caption{\label{fig:3D}Cut through a polymer crystal which was produced by the growth of 
twenty successive layers on a surface with a uniform thickness of 50 units at $T=2.0\,\epsilon k^{-1}$.  
The stems are represented by vertical cuboids. The cut is 16 stems wide. 
}
\end{minipage}
\end{center}
\end{figure}

An understanding of how the thickness of a polymer crystal is determined can be gained from these results
but to do so it is better to plot the thickness of the new layer against the thickness of the growth surface at 
constant temperature, as in Figure \ref{fig:thick.l}.
In our KMC simulations we do not allow the crystalline layer to anneal. 
However, it is likely that some annealing does take place before the next layer is deposited.
In what follows we consider two limits. The first is complete annealing. 
In that case we may replace the previous crystalline layer by a flat substrate with the same thickness. 
The second limit is no annealing. This we discuss with reference to Figure \ref{fig:3D}.
Let us first focus on the complete annealing limit. 
In that case we can use
Figure \ref{fig:thick.l} to consider the effects of growing successive layers on top of each other. 
Following the dotted lines shows one what would happen for growth on an initial surface 50 units thick.  
For example, at $T=2.75\epsilon k^{-1}$ the first layer is 36 units thick, the second 28, the third 23, \dots
The thickness converges to the value $l^{**}$ at which the curve crosses $y=x$ ($l^{**}$=21 at $T=2.75\,\epsilon k^{-1}$), 
i.e.\ to the point where the thickness of the new layer is the same as the previous,
and then the crystal continues to grow at that thickness.
Figure \ref{fig:thick.l}a is another example of a fixed-point attractor. 

\begin{figure}
\begin{center}
\epsfig{figure=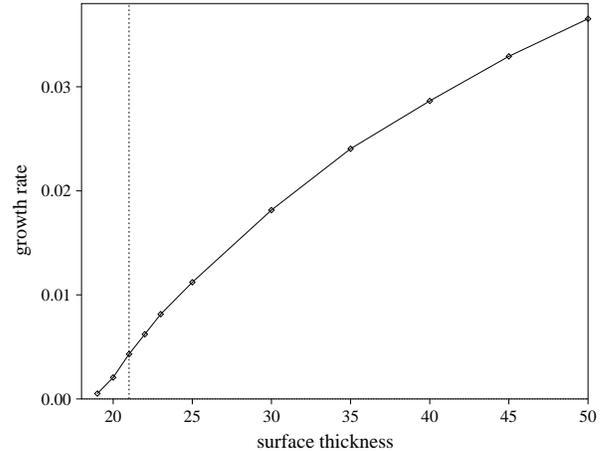,width=8.2cm}
\vglue 0.1cm
\begin{minipage}{8.5cm}
\caption{\label{fig:rate} 
The growth rate (polymer units per $\exp(-\Delta/kT)$ units of time) of the new layer as a function of the thickness of the
underlying surface for $T=2.75\,\epsilon k^{-1}$. A dotted vertical line has been placed at the thickness, $l^{**}$.}
\end{minipage}
\end{center}
\end{figure}

Again this picture is very different from that presented by the LH theory. 
It shows that it is inappropriate to compare the growth rates of crystals of different
thickness because there is only one dynamically-stable thickness, $l^{**}$, for which growth at 
constant thickness can occur. 
For crystals which initially have a thickness different from $l^{**}$,
during growth the thickness will converge to $l^{**}$, as has been observed in experiment.\cite{Bassett62,Dosiere86a}
It is also interesting to note that the growth rate at $l^{**}$ is not a maximum .
Quite on the contrary, the growth of a thick crystal slows as the thickness decreases 
to $l^{**}$ (Figure \ref{fig:rate}).

\end{multicols}
\begin{figure}
\begin{center}
\epsfig{figure=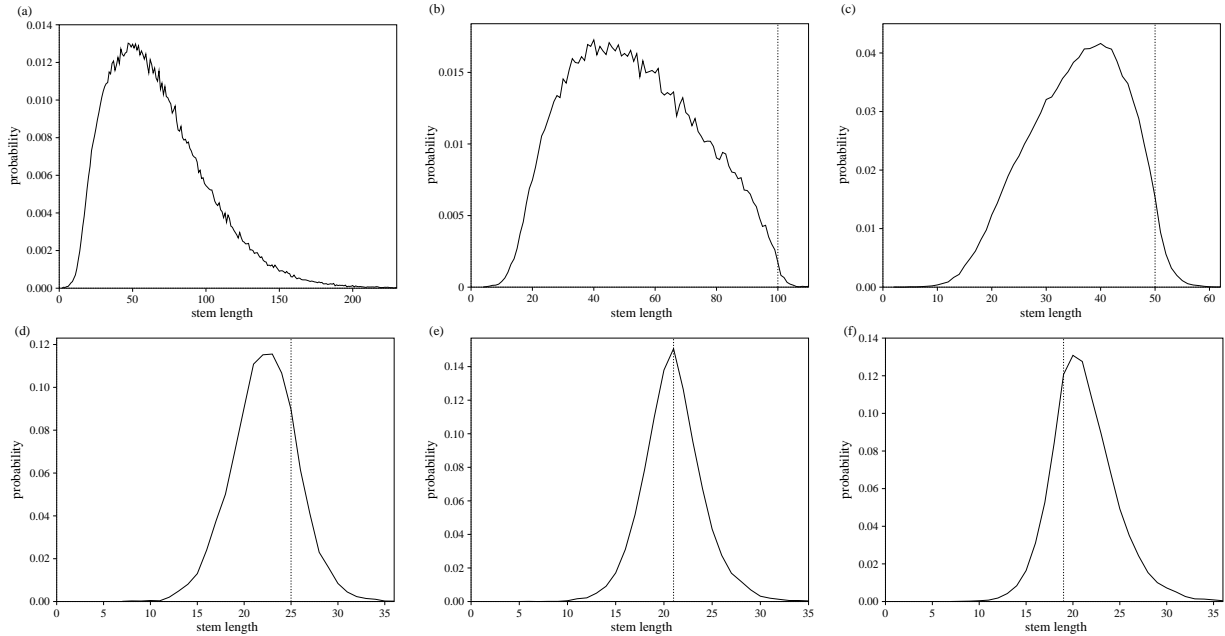,width=16.5cm}
\vglue0.1cm
\begin{minipage}{17.6cm}
\caption{\label{fig:pstem}
Probability distributions
of the stem length for a new crystalline layer grown at $T=2.75\,\epsilon k^{-1}$ 
on a surface of thickness: (a) $\infty$, (b) 100, (c) 50, (d) 25, (e) 21 and (f) 19. 
The dashed vertical lines in the probability distributions are 
at the thickness of the underlying surface.}
\end{minipage}
\end{center}
\end{figure}
\begin{multicols}{2}

Although this viewpoint contrasts with the LH theory,
it was realized in a number  of the early polymer crystallization 
papers\cite{Frank60,Lauritzen67b,Price61} that stable growth could only occur 
at the thickness for which a new layer has the same thickness as the previous.
However, since then this insight has for the most part been neglected,
and, to the best of our knowledge, the iterative maps that underlie it 
have not been previously visualized. 
Also, in recent simulations by Chen and Higgs the lamellar thickness was observed to 
converge to the same value irrespective of the thickness of the initial seed crystal,\cite{Chen98}
but the significance of this behaviour was not fully realized.

We can better understand the reasons for this behaviour by examining representative 
polymer configurations (Figure \ref{fig:config}) and the probability distributions 
of the stem length (Figure \ref{fig:pstem}) for the growth of a single layer at a number of surface thicknesses. 
$l_{\rm min}$ places one constraint on the stem length; only a small
fraction of the stems can be shorter than $l_{\rm min}$ 
if the layer is to be thermodynamically stable. 
The boundary of the growth face places the second constraint on the stem length; 
it is energetically unfavourable for the polymer to extend beyond the edges of the underlying surface. 
Even in the absence of this constraint, 
i.e.\ on the infinite surface, the stem length remains finite because at every step 
there is always a finite probability that a fold will be formed,\cite{Point79a,DiMarzio82a} 
and so the probability distribution decays to zero at large stem lengths (Figure \ref{fig:pstem}a).
For the infinite surface any stem that is longer than $l_{\rm min}$ is stable with respect to the coil and
so a very wide range of stem lengths is observed (the standard deviation in the stem length is $\sim$35 at 
$T=2.75\epsilon k^{-1}$).

As can be clearly seen in Figures \ref{fig:config} and \ref{fig:pstem}, 
the thickness of the growth face exerts an increasingly strong influence on the new layer
as the thickness decreases.
For a surface 100-units thick the probability distribution of the stem length
is very similar to the infinite case---again the maximum occurs at 
$\sim$40--50---except that the long tail of the distribution is cut off at 100
because the polymer rarely extends beyond the edge of the growth face. 
At this thickness the probability of the stem length being greater than the surface thickness is much less
than it being smaller and therefore, the new layer is significantly thinner than the underlying surface. 
As the surface thickness decreases the probability distributions of the stem length 
becomes increasingly narrow and the difference in probability between the stem length being greater
or less than the surface thickness diminishes.
This is because the range of viable stems---those with lengths larger than $l_{\rm min}$ but
which do not significantly overhang the edge of the surface---become smaller. 
Finally, at $l^{**}$ as the surface thickness approaches $l_{\rm min}$ the probability distribution
become symmetrical about the surface thickness and the thickness of the new layer becomes
equal to the thickness of the growth surface (Figure \ref{fig:pstem}e). 
For these reasons the thickness at which stable growth occurs is close to $l_{\rm min}$.
For thicknesses less than $l^{**}$ the asymmetry of the probability distribution is reversed (Figure \ref{fig:pstem}f).

\begin{figure}
\begin{center}
\epsfig{figure=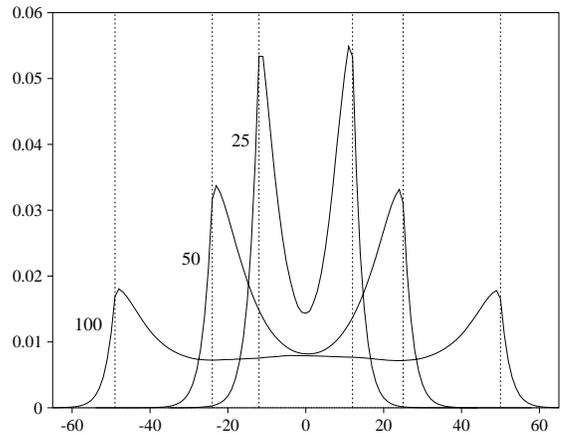,width=8.2cm}
\vglue 0.1cm
\begin{minipage}{8.5cm}
\caption{\label{fig:px} 
The probability distribution for the position of the ends of the crystalline portion of the 
polymer for growth of a single layer on surfaces of different thickness, as labelled. 
The centre of the surface is at 0, and dotted lines have been included at the positions
of the boundaries of the surface.
$T=2.75\,\epsilon k^{-1}$. 
}
\end{minipage}
\end{center}
\end{figure}

It is also possible to understand why the dependence of the growth rate on thickness (Figure \ref{fig:rate})
differs from the predictions of the LH theory. 
The cause is not the decrease of the thermodynamic 
driving force as the thickness decreases (this 
causes the growth rate to go to zero at $l_{\rm min}$) since it is common to both this model and the LH theory.
Rather, the difference arises mainly from the LH assumption that complete stems are deposited in a single step.
Based upon comparisons of this approach with a fine-grained kinetic description of the deposition of a stem,
Frank and Tosi showed that the LH complete-stem approach is reasonable if the only 
point on the fine-grained pathway at which a fold can occur is once a stem is complete. 
However, as has been pointed out previously, this condition does not hold for more realistic 
multi-pathway approaches.\cite{theorynoteb}
That the polymer can form a fold at any point and can overhang the edge of the crystal 
in our model has a considerable effect on the growth rates. 
Such processes can lead to configurations that are not thermodynamically viable, 
such as when a new stem starts to form even though the previous is less than $l_{\rm min}$ 
or when a stem significantly overhangs the edge of the crystal. 
These configurations have to be removed before growth can continue.
The effect of such `blind alleys' is to reduce the growth rate.
As the thickness of the surface decreases the range of stem lengths which give 
viable configurations decreases and so an increasing amount of time is spent 
searching blind alleys. 
These effects are reflected in the probability distributions for the position
of the end of the crystalline polymer (Figure \ref{fig:px}). 
As the thickness decreases the end spends an increasing
proportion of the time near the edge of the growth surface waiting for the polymer to fold at a point
when the stem has a length that is viable for further growth. 

At $T=2.75\,\epsilon k^{-1}$ there is no barrier to growth due to a rounded crystal profile.
When the thickness is $l^{**}$ the new layer has the same thickness as the previous layer.
However, this scenario does not hold for all temperatures.
At any temperature there is a thickness below which
a new layer cannot grow because the surface is too thin,
(e.g. the last point of the curve in Figure \ref{fig:thick.l}a is for a surface 19 units thick)
and there is no {\it a priori\/} reason why this must occur after the thickness curve has crossed $y=x$.
Indeed, for $T>3.2\,\epsilon k^{-1}$ there is no thickness for which successive layers
have the same thickness.
For example, at $T=3.375\,\epsilon k^{-1}$ after the growth of two layers on a 50-unit thick surface
the outer layer is $\sim$29 units thick (Figure \ref{fig:thick.l}b); 
the crystal then stops growing because the outer layer is too thin.
For these smaller supercoolings, as suggested in the entropic
barrier model, the rounding of the crystal profile inhibits growth.

To overcome this barrier requires a cooperative mechanism
whereby a new layer takes advantage of (and then locks in)
dynamic fluctuations in the outer layer to larger thickness.
The presence of such fluctuations is shown in the probability distributions of
Figure \ref{fig:pstem};
however, growth stops in our model because we attempt
to grow a new layer on an outer layer that is static.
As overcoming the barrier would be most rapid when the magnitude of the fluctuations is the minimum necessary,
we expect that this mechanism would lead to the crystal continuing to grow
with the smallest thickness for which a new layer can grow,
e.g. at $T=3.375\,\epsilon k^{-1}$ this is a thickness of 30 (Figure \ref{fig:thick.l}b).
This mechanism again leads to a thickness for the polymer crystal which is close to $l_{\rm min}$.

It is interesting to note that behaviour similar to our results has
been found in some of the extensions of the LH theory which 
include multiple pathways in some restricted or approximate manner.
In a typically prescient paper Frank and Tosi found that growth 
ceases at low supercoolings;\cite{Frank60} however, without a mechanism to allow
growth to continue, they concluded that polymer crystallization experiments
could not correspond to this temperature range.
Lauritzen and Passaglia were also aware that growth 
would cease in their model at low supercoolings, so they introduced an {\it ad hoc\/} 
energetic term into their rate constants that prevented this.\cite{Lauritzen67b,adhoc}

In using Figure \ref{fig:thick.l} to consider the growth of successive layers on the
initial growth face we are assuming that all the roughness in the position of the fold surface
is annealed out before a new layer starts to form. 
To check that the behaviour we observe is not dependent on this assumption we have also performed
simulations where we grow new layers directly on top of the previous grown layers. 
A typical cut through a crystal that was produced by this method is depicted in Figure \ref{fig:3D}. 
We see the same mechanism of thickness determination as before: 
within 5--10 layers the thickness of the crystal converges 
to $l^{**}$ and then growth continues at that thickness.

\begin{figure}
\begin{center}
\epsfig{figure=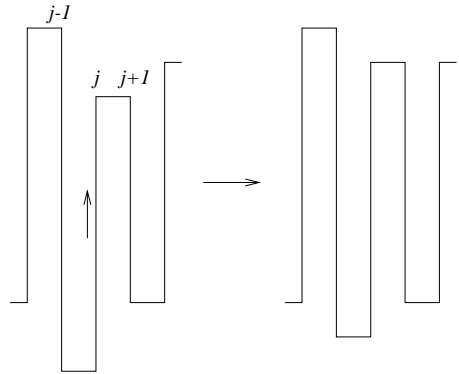,width=6cm}
\vglue 0.1cm
\begin{minipage}{8.5cm}
\caption{\label{fig:anneal} An example of a mechanism by which the roughness in the 
crystalline layer could be annealed. The stem $j$ slides upwards decreasing the length
of stem $j-1$ by one and increasing the length of stem $j+1$ by one. 
The change in energy for this rearrangement is $-\epsilon$.
}
\end{minipage}
\end{center}
\end{figure}

It is not clear which of these two limiting approaches is more realistic. 
In an atomic-force microscopy (AFM) study of the fold surface of polyethylene only on a minority of crystals could
the folds be resolved clearly,\cite{Patil94,Magonov97} suggesting that for most crystals the fold surface retains some roughness.
However, the crystalline configurations generated by our model when the thickness of the growth face 
is much larger than $l_{\rm min}$ (Figure \ref{fig:config}) are clearly too rough. 
One would imagine that local rearrangements such as the one depicted in Figure \ref{fig:anneal} would
act to reduce the roughness whilst keeping the average stem length in the layer constant. 
More uniform configurations could have been obtained by including these mechanisms in our model
but we did not do so because we would have to have set the relative rates of growth and 
annealing to an arbitrary value.
To study the multi-layer growth version of our model more systematically we would probably 
need to include these mechanisms because the roughness can at times hinder the growth; 
e.g.\ it can be hard for a new layer to grow across a section of the previous layer
that involves a particularly short stem.
Chen and Higgs experienced similar difficulties in their simulations.\cite{Chen98}

\begin{figure}
\begin{center}
\epsfig{figure=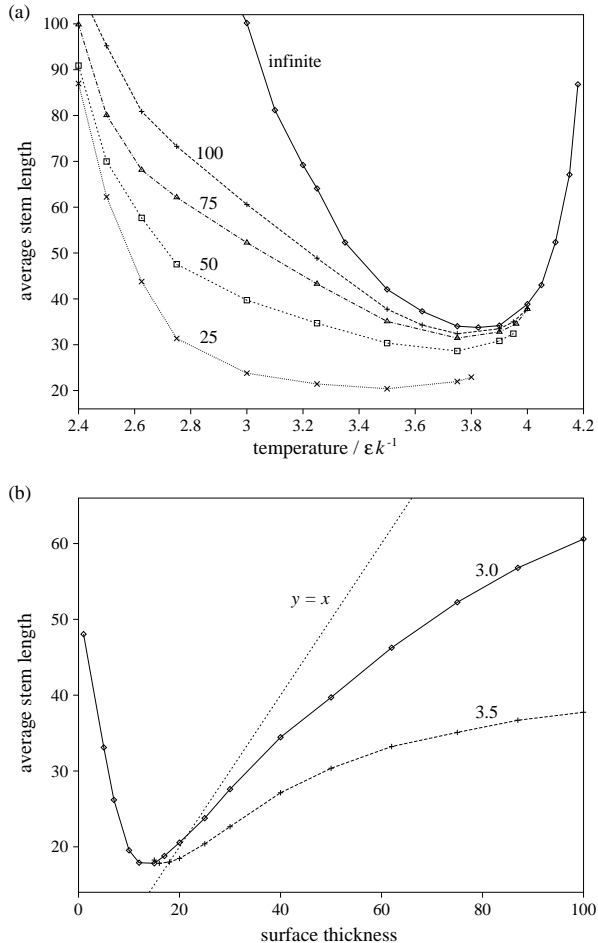,width=8.2cm}
\vglue 0.1cm
\begin{minipage}{8.5cm}
\caption{\label{fig:thick.3D} The dependence of the average stem length in the new crystalline 
layer on the temperature and thickness of the surface when the disordered polymer is assumed to be 
an ideal three-dimensional coil. 
In (a) the labels give the value of the thickness of the underlying surface, 
and in (b) the value of the temperature in $\epsilon k^{-1}$.
}
\end{minipage}
\end{center}
\end{figure}

To determine the effect of our assumption that the disordered polymer
is adsorbed onto the surface we have also performed simulations where the
coil is assumed to be three-dimensional. 
Figure \ref{fig:thick.3D}a shows the dependence of the thickness of a new 
layer on the temperature and thickness of the surface.
The figure has a similar form to Figure \ref{fig:thick.T}a except that 
the rise in the layer thickness at low temperature is not checked by 
the thickness of the underlying layer. 
The free energy difference between the coil and the crystal at low temperature is now large enough 
to allow the polymer to extend beyond the edge of the growth face. 
This has a corresponding effect on $l^{**}$.
When the coil was assumed to be two-dimensional $l^{**}$ was a monotonically increasing
function of temperature which was always slightly larger than $l_{\rm min}$. 
However, now the value of $l^{**}$ increases at low temperatures;
e.g.\ at $T=3.0\,\epsilon k^{-1}$ $l^{**}=21$ whereas at $T=3.5\,\epsilon k^{-1}$ 
$l^{**}=18$ (Figure \ref{fig:thick.3D}b).
Assuming the coil to be three-dimensional does not prevent 
the rounding of the crystal profile which causes growth to stop in
our model at low supercoolings; however, the range of temperature for which this
scenario holds is reduced ($T > 3.8\,\epsilon k^{-1}$) because of the 
greater thermodynamic driving force for crystallization. 

\section{Conclusions}

In summary, our results present a new picture of the mechanisms that
cause the thickness of lamellar polymer crystals to be constrained to a value just above
$l_{\rm min}$.
The free energetic costs of the polymer extending beyond the edges of
the previous crystalline layer and of a stem being shorter than $l_{\rm min}$
provide upper and lower constraints on the length of stems in a new layer.
Their combined effect is to cause
the crystal thickness to converge to a value close to $l_{\rm min}$ where
growth with constant thickness then occurs.

This convergence of the thickness has been observed in
experiments in which the thickness of growing
polymer crystals adjusts to a change in temperature \cite{Bassett62,Dosiere86a} and in
which lamellae form by epitaxial crystallization onto extended-chain
crystalline fibres \cite{Keller79}. It would be very
interesting if AFM could be used to probe the profiles of the steps on 
the lamellae that result from temperature changes.
From these profiles it would be possible to construct maps similar to Fig.\ \ref{fig:thick.l}a
which could confirm the crucial role of the thickness of the
growth face on the properties of a new crystalline layer.
AFM could be also used to study the profile of the crystal close to the growth face
to examine whether rounding of the crystal edge occurs.

Our results have significant implications for the LH surface nucleation theory.
Although in some ways our model is in the LH tradition---new layers nucleate and grow by the addition
of a succession of stems along the growth face---the removal of
many of the LH constraints leads to a significantly different picture for the mechanism 
that determines the thickness of polymer crystals.
Our work undermines many of the assumptions that are crucial to the LH theory:
the initial nucleus is not a single stem; 
the initial nucleus does not determine the thickness within a layer;
the thickness of a new layer is not necessarily the same as the previous layer;
the observed thickness corresponds to the one value of the thickness, $l^{**}$, 
for which crystals can grow with constant thickness and not to the thickness of those 
crystals which grow most rapidly in a fictitious ensemble of crystals of different 
thickness that grow with constant thickness;
the growth rates are significantly affected by `blind alleys' (not included in the LH theory)
which lead to configurations that are not viable for further growth. 
Of course, the attractive feature of the LH theory is that it provides analytical equations
for the thickness and growth rate of polymer crystals that are in reasonable agreement with 
experiment. 
However, our results clearly indicate that the LH equations should be viewed as no more 
than phenomenological because their derivation relies upon
an incorrect description of the physics of polymer crystallization.

The picture of polymer crystallization that comes from our results has more in common
with the entropic barrier approach, and these two models are, in some ways, complementary.
In the simulations of Sadler and Gilmer a realistic description of the polymer connectivity and folds was
sacrificed in order to be able to study the full three-dimensionality of the problem, 
whereas here we have a more realistic description of the polymer but could not at the same time include
the cooperative interlayer dynamics that are necessary for growth at low supercoolings. 
Although Sadler and Gilmer's simulations have been {\it interpreted} in terms of the effects of a
competition between an entropic barrier and a thermodynamic driving force on the growth rate,
we expect that the mechanism of thickness determination is in fact similar to that for our model, namely
that during growth the crystal thickness converges to a thickness $l^{**}$.\cite{SGreex}
The cessation of growth that we see in our model at low supercoolings confirms that the rounding of the
crystal profile does play a role in polymer crystallization.
However, in our model this effect only occurs for a limited temperature range,
whereas in the Sadler-Gilmer model rounding of the crystal profile occurs for all supercoolings. 

We believe that the results in this paper provide new insights into polymer crystallization,
especially into the mechanism by which the thickness of a polymer crystal is attained.
However, we stress that the model we use is simplified in order 
to make it computationally tractable. 
For example, we assume that nucleation only occurs once within a particular polymer; however,
some simulations of a single polymer have seen crystalline domains develop within different parts
of the polymer which at a later time coalesce to form a single crystal.\cite{Yamamoto97,Toma98}
Furthermore, we only allow growth processes to occur within the plane of 
the growth face and only within the outer layer; as explained earlier, the cessation of growth at 
low supercoolings is a result of these simplifications.
This study, then, represents only a starting point in the application of computer simulation
to understand polymer crystallization.
It is hoped that simulation will lead to an increasingly refined picture of the microscopic
mechanisms involved in the growth of polymer crystals. 

\acknowledgements
The work of the FOM Institute is part of the research program of
`Stichting Fundamenteel Onderzoek der Materie' (FOM)
and is supported by NWO (`Nederlandse Organisatie voor Wetenschappelijk Onderzoek'). 
JPKD acknowledges the financial support provided by the Computational Materials Science 
program of the NWO.

\end{multicols}

\begin{thebibliography}{10}
\bibitem{notfree}
The probability distribution of the stem length in the outermost layer has been
  taken as providing evidence of a free energy
  barrier.\cite{Sadler84a,Sadler88a} However, this probability distribution is
  a steady-state distribution under conditions of growth, whereas the free
  energy can only be related to equilibrium probability distributions.
\bibitem{stiff}
The main effect of $\epsilon_g$ is through its effect on $l_{\rm min}$
  (Equations \ref{eq:lmin.2D} and \ref{eq:lmin.3D}).
\bibitem{adhoc}
Lauritzen and Passaglia introduced an energetic penalty if a stem length, $l$,
  differed from the thickness of the surface, $l_s$, which was proportional to
  $|l-l_s|$. If $l>l_s$ this is reasonable since the polymer no longer
  interacts with the surface. However, there is no physical basis for this term
  when $l<l_s$; it just artificially constrains $l$ to have a value similar to
  $l_s$.
\end{thebibliography}
\end{document}